\documentclass[preprint,10pt]{elsarticle}

\makeatletter
\def\ps@pprintTitle{%
\let\@oddhead\@empty
\let\@evenhead\@empty
\def\@oddfoot{\centerline{\thepage}}%
\let\@evenfoot\@oddfoot}
\makeatother

\journal{International Journal of Hydrogen Energy}

\usepackage{graphicx,bbm,setspace,amssymb,amsfonts,amsmath,mathtools,mathrsfs,stmaryrd,amsthm,bm,etoolbox,comment,hyperref,url,caption,subcaption,color,enumitem,algorithm,booktabs,microtype,nicefrac,lingmacros,nccmath,tree-dvips,tikz-cd}
\usepackage[noend]{algpseudocode}
\usepackage{algorithm}

\usepackage[colorinlistoftodos]{todonotes}
\usepackage[utf8]{inputenc} 
\usepackage[T1]{fontenc}    

\usepackage[noend]{algpseudocode}

\newtheorem{theorem}{Theorem}[section]

\newtheorem{thm-defn}[theorem]{Theorem/Definition}

\theoremstyle{definition}

\theoremstyle{remark}
\newtheorem{remark}[theorem]{Remark}

\newcommand{\ignore}[1]{}{}

\begin{document}

\begin{frontmatter}

\title{Spatio-temporal trends in the propagation and capacity of low-carbon hydrogen projects}
   
\author[label1]{Nick James} 
\author[label2]{Max Menzies} \ead{max.menzies@alumni.harvard.edu}
\address[label1]{School of Mathematics and Statistics, University of Melbourne, Victoria 3010, Australia}
\address[label2]{Beijing Institute of Mathematical Sciences and Applications, Tsinghua University, Beijing 101408, China}

\begin{abstract}
This paper uses established and recently introduced methods from the applied mathematics and statistics literature to study trends in the propagation and capacity of low-carbon hydrogen projects over the past two decades. First, we judiciously apply a regression model to estimate the association between various predictors and the capacity of global hydrogen projects. Next, we turn to the geographic propagation of low-carbon hydrogen projects, where we apply a recently introduced method to explore the geographic variance of hydrogen projects over time. Then, we demonstrate that most geographic regions display linear growth in cumulative plants and apply distance correlation to determine the nonlinear dependence between the two most prolific regions - North America and Europe. Finally, we study relationships between the propagation and capacity of green vs blue hydrogen over time, specifically, the time-varying regional consistency between the contribution of green vs blue plants to the total number and capacity of hydrogen plants.
\end{abstract}


\end{frontmatter}

\section{Introduction}

Hydrogen has great potential as an alternative fuel source and may play a role in the world's coordinated attempt to reach net-zero carbon emissions during this century. Hydrogen could be used for numerous purposes, including industrial applications, transport, heating and energy production. By mass, hydrogen contains more than twice the energy potential of natural gas. However, hydrogen in its pure form ($H_2$ gas) does not exist naturally on earth. It must be synthesised via a variety of different procedures.

The production of hydrogen is classified by colours according to its mode of preparation. ``Green hydrogen'' refers to production techniques that do not generate any greenhouse gas (GHG) emissions. This is, of course, the ideal method of production to forward the goal of an alternative energy source that does not contribute further to global warming. Typically, green hydrogen plants use renewable sources of energy (such as solar) to extract hydrogen via the electrolysis of water. This is a desirable process, however the energy production capacity of green plants has been quite limited. On the other hand, black, brown and grey hydrogen refer to production techniques that use black coal, brown coal and natural gas respectively, generating harmful gases including carbon dioxide (CO$_2$) and monoxide. Blue hydrogen is defined as the generation using natural gas, followed by carbon capture and storage (CSS). This is not a truly zero-emissions process, as not all generated GHGs can be captured.

There is significant variability in the countries' level of sophistication with regards to alternative and clean energy. Many European countries such as Germany have been able to reduce carbon dioxide emissions over the past 50 years, primarily driven by their willingness to drive the adoption of new technologies, often with significant initial cost. Less willing to wear the immediate economic consequences, many developing countries, as well as the United States and Australia, have been more hesitant to commit to green and alternative energy sources. Instead, they continue to utilise energy sources that have already reached economies of scale.

With the increase in global interest towards alternative energy sources in recent years, a great deal of research has focused on the viability and efficiency of hydrogen production. Early research into the efficiency of hydrogen fuel cells dates back several decades \cite{doenitz_hydrogen_1980,wendt_hydrogen_1991,khaselev_high-efficiency_2001}. Since then, research on hydrogen production has taken numerous directions. \cite{ursua_hydrogen_2012} and \cite{chi_water_2018} each provide a review article on different forms of electrolysis. \cite{zeng_recent_2010} provides a detailed examination of alkaline electrolysis (ALK), while \cite{barbir_pem_2005,grigoriev_pure_2006,bessarabov_pem_2016} survey successive advances in proton exchange membrane electrolysis (PEM). \cite{guo_comparison_2019} compares advantages and disadvantages of ALK and PEM in detail. 

More recent research has explored several different means to enhance efficiency of hydrogen production and make use of different renewable energy sources. \cite{yilmaz_thermodynamic_2014} explores the use of geothermic energy to power hydrogen electrolysis; \cite{rozendal_principle_2006} investigates the use of micro-organisms (biocatalysed electrolysis). \cite{call_hydrogen_2008} discusses advances in such microbial electrolysis cells. More recently, \cite{Rai2022} outlines how electronic waste may be utilised to generate metallic components for a process called ``chemical looping reforming.''

Regarding increased efficiency, there have been many technological advances, both novel and incremental. \cite{wang_intensification_2014} examines means to reduce energy consumption during electrolysis; \cite{chakik_effect_2017} discusses the different efficiency of different electrodes, with further advances explored by \cite{Wu2022}. In particular, \cite{Ikeda2022} analyses electrode overpotential during ALK electrolysis, while \cite{zhang_evaluation_2010} explores optimal configurations of electrolysis under several different conditions. \cite{Shit2022} investigates improved and modern electrocatalysts, \cite{Tong2022} analyses the use of hybrid structures for highly efficient water electrolysis, combining both morphological features and electrochemical properties, while various researchers have studied cutting-edge catalysts \cite{Liang2022,Khasanah2022,Liu2022_hydrogen}.

In addition, numerous articles discuss the potential for hydrogen production in one region at a time (usually a single country), both in reference to its natural resources and policy environment. Analysis has covered countries as diverse as the USA \cite{lattin_transition_2007}, China \cite{yuan_hydrogen_2010}, Mexico \cite{ramirez-salgado_roadmap_2004}, Argentina \cite{sigal_assessment_2014}, the United Kingdom \cite{park_country-dependent_2013}, South Korea \cite{park_country-dependent_2013}, Morocco \cite{touili_technical_2018}, the Philippines \cite{collera_opportunities_2021}, and various countries throughout the European Union \cite{apak_renewable_2012}.

Whereas the aforementioned existing papers have been more technological or geopolitical in focus, our paper is a mathematical study of trends in the rollout and prevalence of low-carbon hydrogen plants by both technology and location. We make heavy use of \emph{time series analysis}, which has been extensively applied to various fields such as epidemiology \cite{James2021_virulence,Manchein2020,Blasius2020,james2021_CovidIndia,Estrada2020,james2021_TVO}, finance \cite{Basalto2007,james2021_mobility,Liu1997,Drod2001,arjun,Drod2018,Drod2019,james2021_MJW,Drod2020}, digital currencies \cite{james2021_crypto2} and even sport \cite{Ribeiro2012,james2021_olympics}. We are unaware of any instance where time series analysis has been applied to the rollout of hydrogen plants over time and by location. We study the changing propagation and capacity of hydrogen plants over time, with a particular interest in the increasing potential of green hydrogen plants, which emit zero carbon. We also investigate differences in these rollouts on a geographic basis. Our main finding is promising: an exponential increase in the capacity of green plants over time and a dramatic closing between the capacity of green and non-green plants.

\section{Data}
\label{sec:data}
Our data comes from the International Energy Agency \cite{Hydrogendata}, and consists of plants built in (or projected to be built in) 2000-2028, a period of $T=29$ years. This dataset records all low-carbon hydrogen plants, namely either green (zero emissions) or blue (incorporating fossil energy and carbon capture and storage). Each plant is classified according to one of five underlying technologies: four different types of water electrolysis, which are all green plants, and ``fossil'', indicating the use of fossil fuels. The four green technologies are alkaline electrolysis (ALK), proton exchange membrane electrolysis (PEM), solid oxide electrolysis cells (SOEC), and other electrolysis. By aggregating the four green electrolysis technologies, the plants are classified as green or non-green (``fossil''). For each plant, the location is also recorded, either specified by the country of location or continent. As such, all our analysis proceeds on a continent-by-continent basis, in which we record and analyse only the continent of location. We divide the plants into continental groups as follows: North America, South America, Europe, Oceania, East Asia (China and Japan) and Other Asia (mostly consisting of Indian plants).

\section{Trends in capacity and propagation of plants}
\label{sec:regression}

\begin{figure}
    \centering
\begin{subfigure}{0.7\textwidth}
    \includegraphics[width=\textwidth]{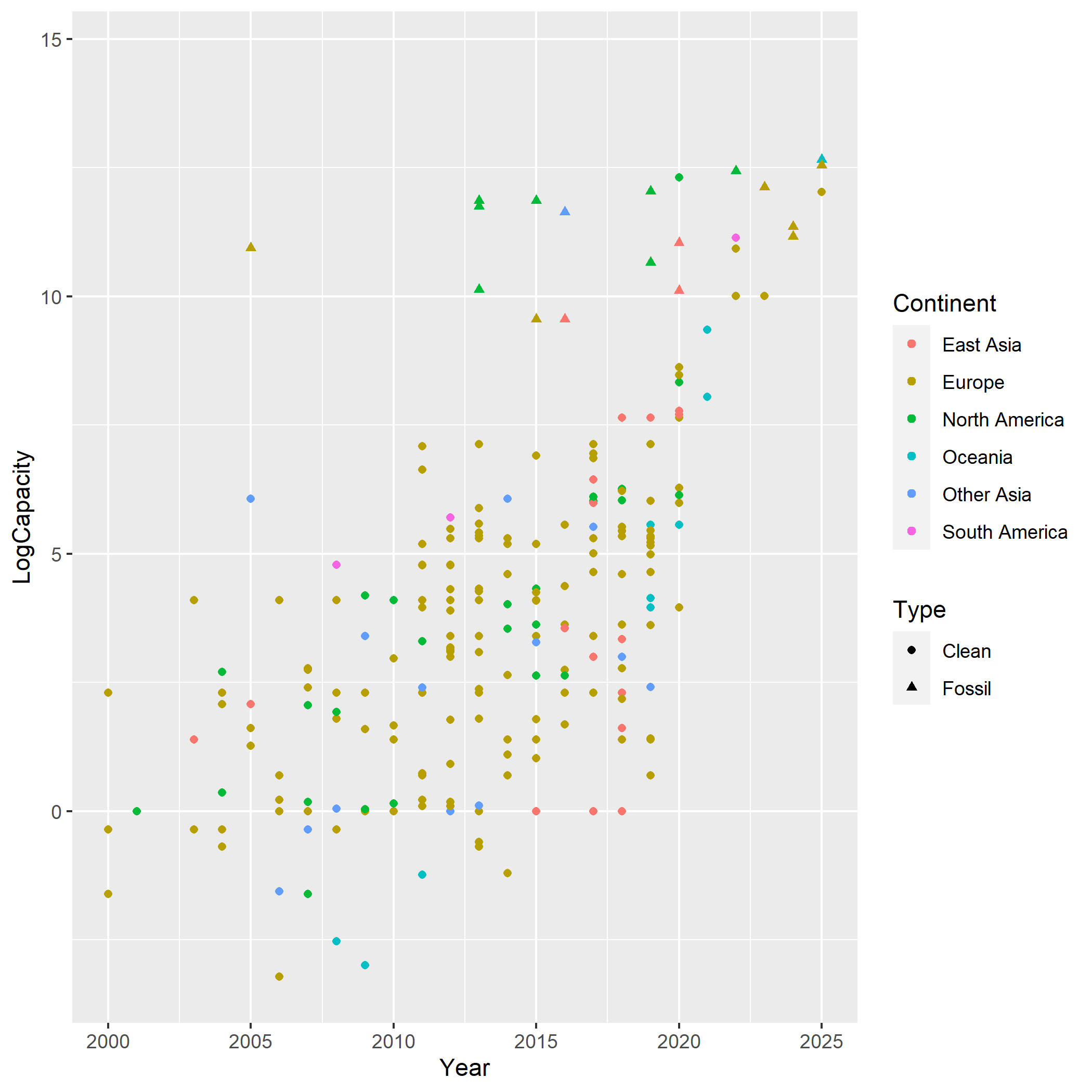}
    \caption{}
    \label{fig:Scatterplot1}
    \end{subfigure}
\begin{subfigure}{0.7\textwidth}
    \includegraphics[width=\textwidth]{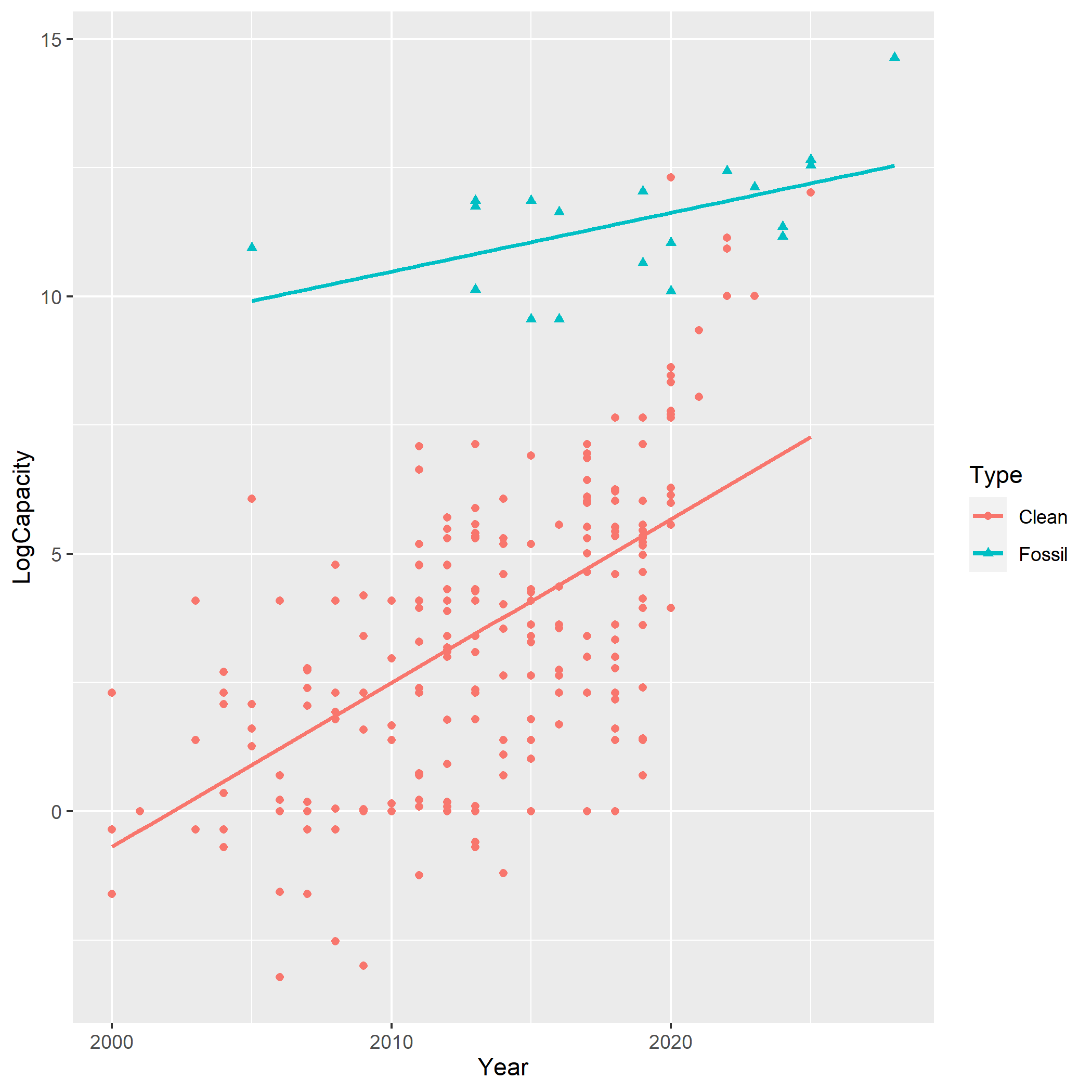}
    \caption{}
    \label{fig:Scatterplot2}
    \end{subfigure}
    \caption{Log capacity vs year of capacity for all plants in our data set with a recorded capacity. In (a) we separate plants by continent of construction, while in (b) we display linear trends for green vs fossil (blue) plants. We can clearly see the large number of European plants featured in (a), and the early dominance of fossil plants by several orders of capacity magnitude in (b). Fortunately, this difference is closing with time.}
    \label{fig:scatterplots}
\end{figure}

In Figure \ref{fig:scatterplots}, we display all the plants with a known energy capacity in our dataset. Displaying the logarithm of the capacity against the year of construction, we see an approximate linear trend between the log of capacity and the year. This suggests an exponential growth in capacity vs time. We show the separation of plants by continent in Figure \ref{fig:Scatterplot1} and by technology in Figure \ref{fig:Scatterplot2}. Earlier on, we see that fossil (blue) plants are several orders of magnitude greater in capacity than green plants, but this difference reduces dramatically over time (\ref{fig:Scatterplot2}). 

To elucidate these findings, we implement two linear regressions:
\begin{align}
\label{eq:linearlinear}
   y_i &= \beta_0 + \beta_1 t_i + \sum_{j=1}^5 \beta_j^{con} \bm{1}_{ij}^{con} + \sum_{k=1}^4 \beta_k^{tech} \bm{1}_{ik}^{tech},\\
   \label{eq:loglinear}
       \log( y_i) &= \beta_0 + \beta_1 t_i + \sum_{j=1}^5 \beta_j^{con} \bm{1}_{ij}^{con} + \sum_{k=1}^4 \beta_k^{tech} \bm{1}_{ik}^{tech} 
\end{align}
We encode six continents as Europe, North America, South America, East Asia (Japan and China), Other Asia (predominantly India), Oceania, with Europe as the default categorical variable. We also consider the five technologies: ALK, PEM, SOEC, other electrolysis and Fossil, with ALK as our default variable. For model (\ref{eq:linearlinear}), the adjusted $R^2$ (a measure of goodness of fit \cite{ISL}) is 0.17, while for (\ref{eq:loglinear}), the adjusted $R^2$ is 0.65. This strongly suggests a better fit where capacity is predicted and understood to increase exponentially over time, confirming the qualitative observations of Figure \ref{fig:scatterplots}.

We include additional details on model (\ref{eq:loglinear}) in Table \ref{tab:linearregression}. Relative to Europe as a baseline, South America has a significantly greater capacity ($p=0.042$, indicating a significant contribution to the linear regression \cite{ISL}), while PEM and SOEC have less capacity than ALK. This could be of potential interest to those interested in owning and operating green hydrogen plants. In particular, one could consider the profitability implications when contrasting the varying costs of electrolysis utilising PEM, SOEC and ALK, and a candidate plant's respective output.

\begin{table}[ht]
\centering
\begin{tabular}{rrr}
  \hline
 & Estimate & p-val \\ 
  \hline
  Year & 0.332284 & $<10^{-16}$ \\ 
  PEM & -1.279447 & 0.000046 \\ 
  SOEC & -2.473728 & 0.000052 \\ 
  Other electrolysis & 0.771256 & 0.174125 \\ 
    Fossil & 5.459489 & $<10^{-16}$ \\ 
  East Asia & -0.774617 & 0.110566 \\ 
  North America & 0.497280 & 0.223130 \\ 
  Oceania & -0.850573 & 0.171181 \\ 
  Other Asia & -1.037974 & 0.067710 \\ 
  South America & 2.455060 & 0.041843 \\ 
  \hline
\end{tabular}
\caption{Log capacity against tech and continent. Default tech is ALK, default continent is Europe. The adjusted $R^2$ is 0.65. As expected, a highly significant $p$-value is observed for fossil plants, which feature several orders of magnitude higher capacity than green plants. Green technologies PEM and SOEC exhibit significantly reduced capacity relative to ALK plants.}
\label{tab:linearregression}
\end{table}

\section{Geographic variance}
\label{sec:geographic}

In this section, we study the geographic propagation of hydrogen plants around the world. We use  same $n=6$ continent groups as Sections \ref{sec:data} and \ref{sec:regression}. We consider multivariate time series $x_i(t)$ and $y_i(t)$ consisting of the number of new plants and total capacity produced thereof, respectively, on a continent-by-continent basis, $i=1,...,n, t=1,...,T.$ We wish to investigate the changing geographic spread of hydrogen plants and their energy production capacity with time. For that purpose, we convert these time series into rolling distributions as follows. Let $f^{[a:b]} \in \mathbb{R}^n$ be the probability vector of new plants in each continent, observed across a period (in years) $a \leq t \leq b$, divided by the total number of new plants in this period:
\begin{align}
f^{[a:b]}_i = \dfrac{\sum_{t=a}^b x_i(t)}{\sum_{t=a}^b \sum_{j=1}^{n} x_j(t)}, i={1},...,n. 
\end{align}
Analogously, let $g^{[a:b]} \in \mathbb{R}^n$ be the probability vector of green plants in each continent. We may also make an analogous definition for all or green plants' distributions of capacity.

\begin{figure}
    \centering
    \begin{subfigure}{0.49\textwidth}
        \includegraphics[width=\textwidth]{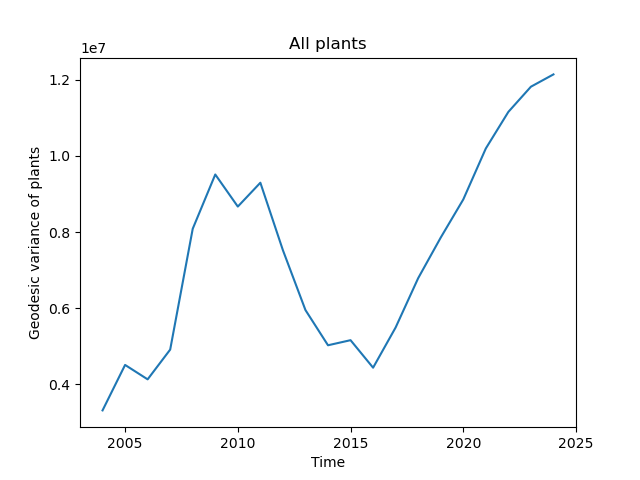}
\caption{}
\label{fig:variance_plants_all}
    \end{subfigure}
 \begin{subfigure}{0.49\textwidth}
        \includegraphics[width=\textwidth]{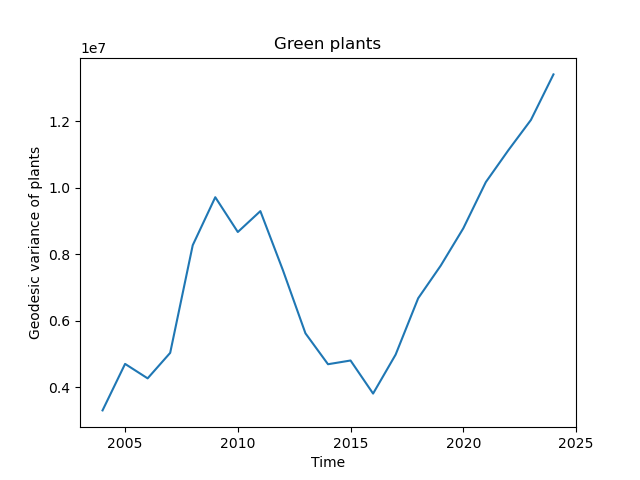}
\caption{}
\label{fig:variance_plants_green}
    \end{subfigure}
\caption{Geodesic variance of (a) all hydrogen plants and (b) green hydrogen plants. The initially low geographic variance is due to the dominance of Europe. Subsequently, the worldwide variance briefly increases and then decreases, as propagation throughout Asia occurs and then drops off. From 2016, the geographic variance begins to rise sharply, due to construction and future plans for hydrogen plants in China and around the world.}
\label{fig:variance}
\end{figure}

We use the notion of geodesic variance developed in \cite{James2021_geodesicWasserstein} to quantify the spread of a probability distribution across a metric space, while taking the spatial structure into account. Given a distribution $f$ corresponding to a measure $\mu$ on a finite metric space $(X,d)$, let
\begin{align}
\label{eq:geodesicvariance}
\text{Var}(f) &= \int_{X \times X} d(x,y)^2 d\mu(x) d\mu(y) \\ &= \sum_{x, y \in X} d(x,y)^2 f(x)f(y),    
\end{align}
where the second equality is valid when the metric space is discrete, as in this example. We term (\ref{eq:geodesicvariance}) the \emph{geodesic variance} of the distribution $f$. In this instance, for any distribution $f$ across continents indexed $i=1,...,n$, the geodesic variance is calculated as
\begin{align}
\label{eq:geodesicvariance2}
\text{Var}(f)  = \sum_{i,j=1}^n d(i,j)^2 f_if_j,    
\end{align}
where $d(i,j)$ is the real-world haversine \cite{haversine} distance between the centroids of continents indexed by $i$ and $j$. In Figure \ref{fig:variance_plants_all}, we plot the time-varying geographic variance of grouped 5-year distributions, $t \mapsto \text{Var}(f^{[t-4: t]}), t=5,...,T$, and analogously for green plants $g$. This captures the geographic spread of new (and specifically green) plants throughout a rolling 5-year period.

Both Figures \ref{fig:variance_plants_all} and \ref{fig:variance_plants_green} exhibit the same trajectory between 2004-2024, where geodesic variance increases until 2008, decreases until 2016, and then increases until 2024. Both figures start with an initially low geodesic variance in 2005, which is likely explained by Europe's dominance in hydrogen plant propagation. During the initial increase, we see several plants appear throughout Asia, which leads to an increase in the variance. Between 2010-2015, the geodesic variance declines, which is likely due to East Asia's levelling off in hydrogen plant propagation. In 2015-2016, East Asia experiences a significant boost in hydrogen plant commencement. This is largely due to China, whose first plant appears in 2017. Since then, China has accounted for numerous other hydrogen plants. We see a similar pattern for our two collections, all hydrogen plants and green hydrogen plants exclusively. This is predominantly due to the green hydrogen plants accounting for the vast majority of all hydrogen plants. The sharp rise in geodesic variance beyond 2016, including future projects, may indicate the increased awareness of clean energy and decarbonisation.

\begin{figure}
    \centering
\begin{subfigure}{0.49\textwidth}
\includegraphics[width=\textwidth]{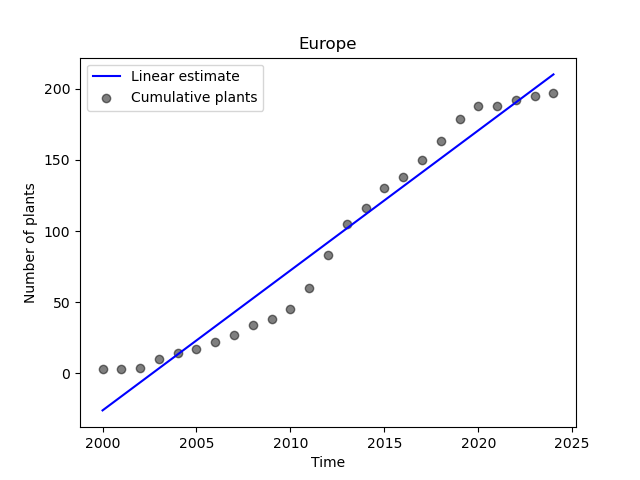}
\caption{}
\label{fig:europe_fit}
\end{subfigure}
 \begin{subfigure}{0.49\textwidth}
\includegraphics[width=\textwidth]{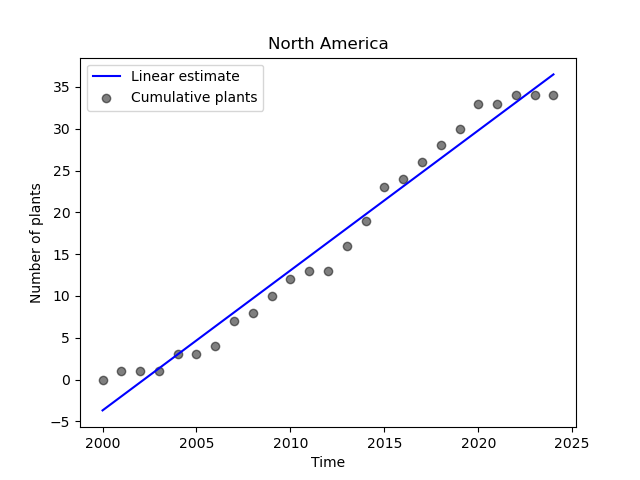}
\caption{}
\label{fig:north_america_fit}
\end{subfigure}
 \begin{subfigure}{0.49\textwidth}
\includegraphics[width=\textwidth]{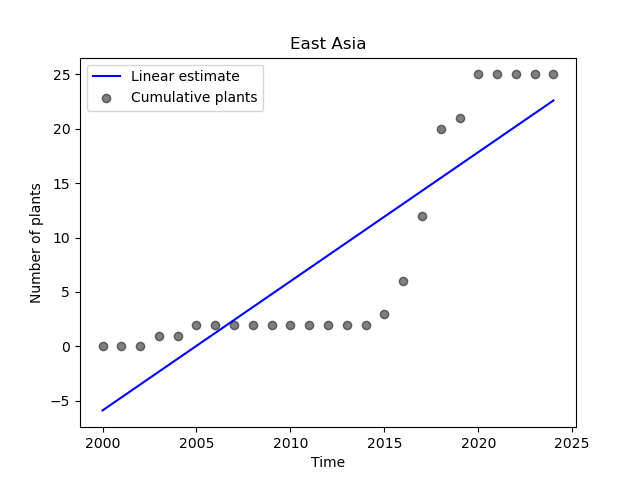}
\caption{}
\label{fig:East_Asia_fit}
\end{subfigure}
\begin{subfigure}{0.49\textwidth}
\includegraphics[width=\textwidth]{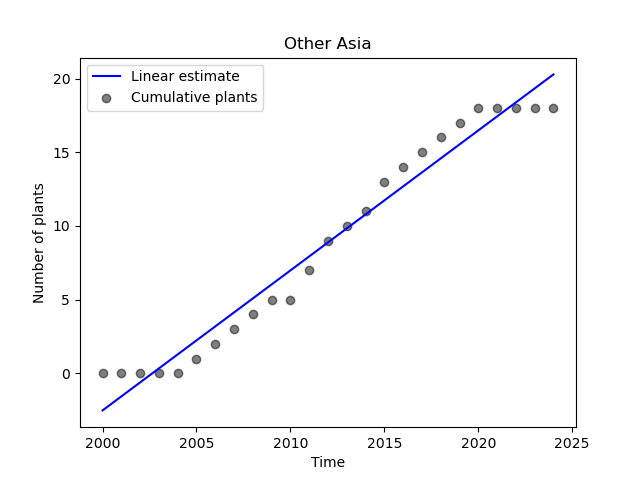}
\caption{}
\label{fig:Other_Asia_fit}
  \end{subfigure}
\caption{Linear estimates on cumulative plants for (a) Europe, (b) North America (c) East Asia (d) Other Asia. Europe, North America and Other Asia display a relatively consistent linear trend, while East Asia is relatively constant between 2000-2015, with subsequent rapid growth.}
\label{fig:linearfits}
\end{figure}

To elucidate the propagation of plants by continent, we plot the cumulative number of plants and fit linear trends to them in Figure \ref{fig:linearfits}. We highlight the results of Europe, North America, East Asia and other Asia in Figures \ref{fig:europe_fit}, \ref{fig:north_america_fit}, \ref{fig:East_Asia_fit} and \ref{fig:Other_Asia_fit}, respectively. Europe, North America and Other Asia all display a relatively consistent linear trend throughout the entire period of analysis. However, East Asia's cumulative number of plants is not modelled well by a linear fit. The cumulative number of plants is perhaps best modelled by a hyperbolic function, given the relatively constant level between 2000-2015, followed by rapid growth between 2015-2020 (mostly due to plants in China), and another constant period during the early-mid 2020's (namely, a lack of planned future projects). These patterns, which demonstrate the consistency of hydrogen plant propagation over time, may indicate the evolution of each continent's interest in decarbonisation and hydrogen production over the last two decades and near future.

\begin{figure}
    \centering
    \includegraphics[width=0.7\textwidth]{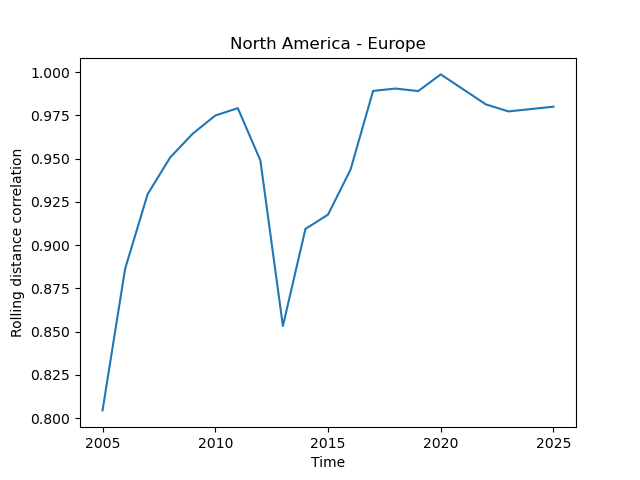}
    \caption{Distance correlation between cumulative number of plants in North America and Europe. A dip in distance correlation around 2013 is indicative of markedly varying concavity between the growth of European and North American plants.}
    \label{fig:distancecorr}
\end{figure}

To complement the above analysis, we explore the rolling distance correlation \cite{Szkely2007} between the cumulative number of plants in North America and Europe. Distance correlation is not to be confused with the better known, more widely used Pearson correlation \cite{ISL}. Distance correlation captures linear and nonlinear associations between two random variables, while Pearson correlation can only detect linear relationships. The distance correlation is computed as follows. Let $p(t)$ and $q(t)$ be the cumulative number of plants as of time $t$ in Europe and North America, respectively, $t=1,...,T$. Let $1 \leq a \leq b \leq T$. We define
\begin{align}
    d_{s,t}=|p(t) - p(s)|;\\
    \bar{d}_{s,\mu}[a:b] = \frac{1}{b-a+1}\sum_{t=a}^{b} d_{s,t}; \\
    \bar{d}_{\mu,t}[a:b]=  \frac{1}{b-a+1}\sum_{s=a}^{b} d_{s,t};\\
    \bar{d}[a:b] = \frac{1}{(b-a+1)^2} \sum_{s,t=a}^{b} d_{s,t};\\
    A_{s,t}[a:b]=d_{s,t} + \bar{d}[a:b]  - \bar{d}_{s,\mu}[a:b] -  \bar{d}_{\mu,t}[a:b].
\end{align}
Analogously, we define $B_{s,t}[a:b]$ in terms of $q(t)$. Then the distance correlation of plants built in the period $a \leq t \leq b$ is defined as follows:
\begin{align}
    DC[a:b]= \frac{\sum_{s,t=a}^{b} A_{s,t}[a:b]B_{s,t}[a:b]}{\left(\sum_{s,t=a}^{b}A_{s,t}[a:b]^2\right)^\frac12\left(\sum_{s,t=a}^{b}B_{s,t}[a:b]^2\right)^\frac12}.
\end{align}
In Figure \ref{fig:distancecorr}, we plot the rolling distance correlation $t \mapsto DC[t-4:t], t=5,...,T.$

Early on in our analysis window, the distance correlation is low - which is predominantly due to the sparsity of data. Beyond that, our time-varying distance correlation exhibits a local minimum in 2013. During this time, Europe and North America display profound differences in their concavity behaviours in new plants. During this period, Europe's cumulative plants display a concave up shape, while North American cumulative plants display a concave down shape. Beyond this point, both North America and Europe exhibit relatively consistent linear increases, which is reflected in the high distance correlation between these two regions' cumulative plants.

\section{Evolutionary relationships between fossil and green hydrogen}
\label{sec:inconsistency}

\begin{figure}
    \centering
 \begin{subfigure}{0.49\textwidth}
        \includegraphics[width=\textwidth]{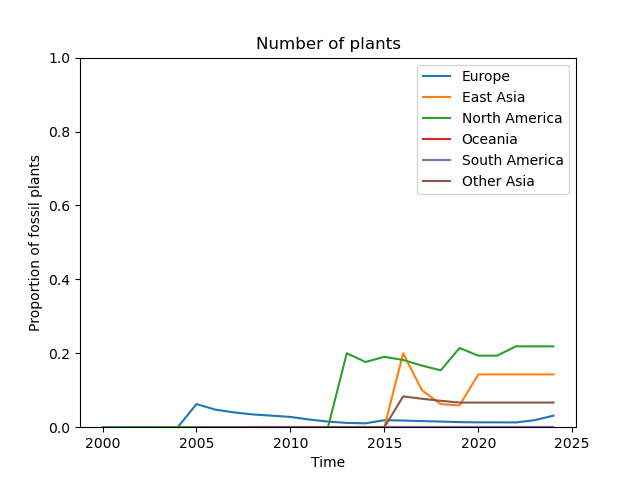}
\caption{}
\label{fig:numberratio}
    \end{subfigure}
    \begin{subfigure}{0.49\textwidth}
        \includegraphics[width=\textwidth]{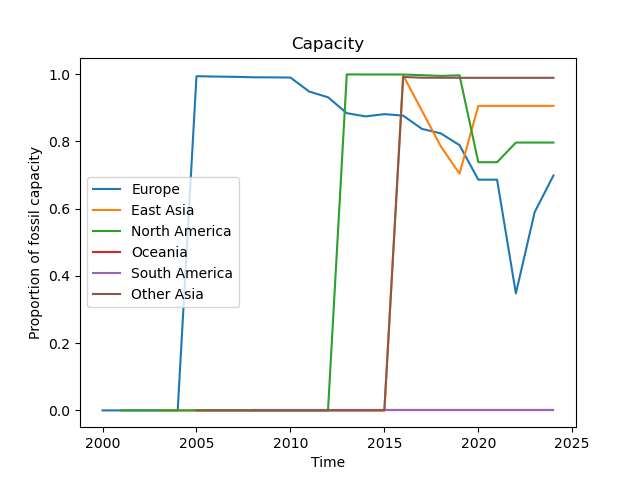}
\caption{}
\label{fig:capacityratio}
    \end{subfigure}
 \begin{subfigure}{0.7\textwidth}
        \includegraphics[width=\textwidth]{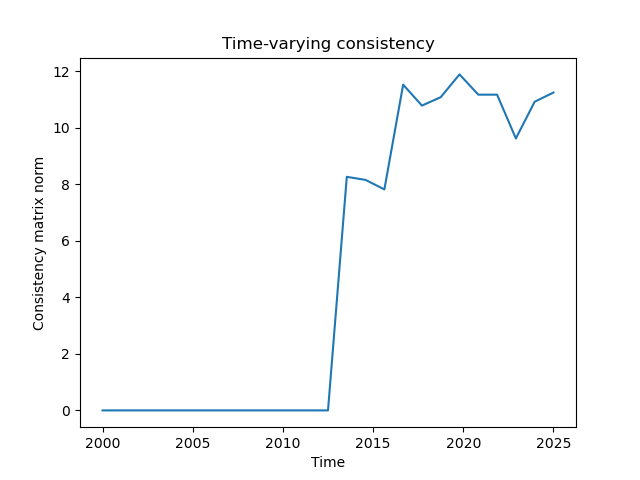}
\caption{}
\label{fig:CCA}
    \end{subfigure}
\caption{In (a) and (b), we plot the proportion of fossil plants by count and capacity in each of the six continents under consideration. In (c), we plot the time-varying inconsistency norm $\nu(t)$, defined in (\ref{eq:nut}). This shows the growing inconsistency between differences in continents' proportion of green vs blue counts of plants and capacity thereof.}
\label{fig:ratios}
\end{figure}

In this section, we study relationships between the contribution of fossil vs green hydrogen plants to the total number of plants and energy capacity they can produce. Specifically, let $p_i(t),c_i(t)$ be the cumulative number of plants and capacity thereof established as of time $t$ in continent $i$, $i=1,...,6$. Of the $p_i(t)$ plants, suppose that the proportion (fraction) of plants that are green and fossil are $p^g_i(t)$ and $p^f_i(t)$ respectively, normalised so that $p^g_i(t)+p^f_i(t)=1$. Suppose that $c^g_i(t)$ and $c^f_i(t)$ are the analogous proportions of total capacity up to time $t$ produced by green and fossil plants, respectively. If there are no plants up to time $t$ in continent $i$, let $p^f_i(t)=c^f_i(t)=0$.

Thus, $(p^g_i(t), p^f_i(t))$ and $(c^g_i(t), c^f_i(t))$ are probability vectors in $\mathbb{R}^2$. For way of example, if a certain continent has only green plants as of time $t$, then $(p^g_i(t), p^f_i(t))=(1,0)$. For each $t$, we form $6 \times 6$ matrices between countries for counts of plants and capacity as follows:
\begin{align}
    D^p_{ij}(t)=| p^f_i(t) - p^f_j(t)|= | p^g_i(t) - p^g_j(t)|= \frac12 \| (p^g_i(t), p^f_i(t))- (p^g_j(t), p^f_j(t)) \|_1; \\
     D^c_{ij}(t)= | c^f_i(t) - c^f_j(t)|=| c^g_i(t) - c^g_j(t)|= \frac12 \| (c^g_i(t), c^f_i(t))- (c^g_j(t), c^f_j(t)) \|_1.
\end{align}
We write the three equivalent forms, particularly the third, to emphasise that this distance can be understood as the \emph{discrete Wasserstein distance} \cite{James2021_geodesicWasserstein} between probability vectors $(p^g_i(t), p^f_i(t))$ and $(p^g_j(t), p^f_j(t))$ (for counts of plants, and analogously for capacity). 
\begin{remark}
The three equivalent definitions of $D^p_{ij}(t)$ are only valid in the case when continent $i$ has a non-zero cumulative number of plants up until time $t$. In the instance where there are no plants, we have set $p^f_i(t)=c^f_i(t)=0$, and then the first definition
\begin{align}
    D^p_{ij}(t)=|p^f_i(t) - p^f_j(t)|
\end{align}
is valid, and similarly for $D^c$.
\end{remark}

To each distance matrix $D$, we associate an affinity matrix $A$ defined by
\begin{align}
    A_{ij}=1-\frac{D_{ij}}{\max D},
\end{align}
where the entries of $A$ are normalised to lie in $[0,1]$. Further, for an $n \times n$ matrix $A$, let its \emph{norm} be defined as $\|A\|=\sum_{i,j=1}^n |a_{ij}|$. This is a measure of the total size of the matrix. Let the affinity matrices associated to $D^p(t)$ and $D^c(t)$ be $A^p(t)$ and $A^c(t)$ respectively. Finally, we define the inconsistency matrix between plants and their capacity:
\begin{align}
    \text{Con}^{p,c}(t)=A^p(t) - A^c(t); \\
    \nu(t)=\| \text{Con}^{p,c}(t) \|
    \label{eq:nut}
\end{align}
The norm of this matrix measures the collective inconsistency between differences in continents' proportion of green vs blue (cumulative) counts of plants and capacity thereof.

In Figures \ref{fig:numberratio} and \ref{fig:capacityratio}, we plot the proportion of fossil plants and capacity, $p_i^f(t)$ and $c_i^f(t)$, respectively, with time. We see that initially, there are no fossil plants, but as soon as the first appears in each continent, it dominates close to 100\% of that continent's hydrogen energy capacity. However, this dips to more reasonable proportions as we approach the end of the period of analysis. This confirms the closing gap between the production capacity of green vs fossil plants observed in Section \ref{sec:regression}.

In Figure \ref{fig:CCA}, we display the time-varying inconsistency matrix norm $\nu(t)$ over our period. This total inconsistency is zero until 2011 due to either the complete lack of fossil plants (hence no inconsistency between the production vs number of fossil vs green plants) and then only Europe having fossil plants. During the latter, the affinity matrices for both capacity and plant count have a block diagonal structure:
\begin{align}
\begin{bmatrix}
1 & 0 & 0 & 0& 0 & 0\\ 
0 & 1 & 1 & 1 & 1 & 1\\
0 & 1&  1& 1 & 1 & 1\\
0 & 1&  1& 1 & 1 & 1\\
0 & 1&  1& 1 & 1 & 1\\
0 & 1 &  1 &1 & 1 & 1
\end{bmatrix}_.
\end{align}
Equivalently, the differences between continents with respect to the blue proportion of plant counts are directly proportional to differences between continents with respect to blue capacity, so we record no inconsistency.

Then in 2013, we see North America generate a fossil project. This constitutes over 99\% of North America's total capacity, but just 20\% of its plants by count. At this time, Europe has 88\% of its total capacity generated by fossil plants, which are 1.2\% by count, exhibiting different proportions than North America. This is first introduction of inconsistency between the continents in terms of their differences between their proportions of fossil and green number of plants vs capacity. Subsequently, when other continents start to roll out fossil-related projects, they introduce differing proportions of capacity vs number of plants that are fossil, promoting further inconsistency in the proportion of fossil projects by count and their capacity. In 2016, both Other Asia and East Asia establish their first fossil plants. This introduces the additional inconsistency observed towards the end of the period. For example, in the year 2021, fossil (blue) plants constitute 1.3\% of Europe's plants by count but 69\% by capacity; East Asia's plants constitute 14\% and 91\% respectively. As a final remark, we see that Europe's proportion of total capacity generated by green plants is consistently rising. This supports the finding of greater competitiveness of green plants with time in Figure \ref{fig:scatterplots}.

\section{Conclusion}

This paper is the first to study the spatio-temporal propagation of hydrogen projects and their associated capacity using a range of techniques from  applied mathematics. Combining the findings from studying the distinct phenomena explored in various sections of this paper provides a unique, holistic overview regarding the state of low-carbon hydrogen projects globally. No similar work has been found in the application field of low-carbon hydrogen.

First, we demonstrate changing relationship between green (zero carbon) and blue (captured and stored carbon) hydrogen project capacity. Early in our analysis window, blue hydrogen projects dominated green projects with respect to their capacity by several orders of magnitude. However, throughout our analysis, we have observed exponential growth in the capacity of green hydrogen projects. This is an excellent sign for the future of this renewable fuel and suggests that investing in green hydrogen projects will likely provide returns in the coming years.

Our methods provide frameworks for mathematically validating this finding in two separate settings. First, in our fitting the log-linear regression model, we demonstrate improved performance (indicated by much higher $R^2$) of our model after taking a log transformation of our response variable. Second, in Section \ref{sec:inconsistency}, Figure  \ref{fig:capacityratio} shows a recent and welcome decrease in the dominance of non-green hydrogen plants' capacity over green. These frameworks provide a greater degree of rigour, and allow us to confirm the hypothesis formulated after our initial exploratory analysis (Figure \ref{fig:scatterplots}).

In Section \ref{sec:geographic}, our application of the geodesic Wasserstein metric and the linear estimates on regional plant numbers tell a coherent story. First, the ``up-down-up'' pattern in the geodesic variance of green hydrogen plants confirms the evolution of global interest in hydrogen energy. The early dominance of European countries, followed by brief propagation to Japan and North America, a subsequent plateau, and then reinvigorated interest worldwide, is reflected in Figure \ref{fig:variance}. The linear estimates of cumulative hydrogen plants applied to Europe, North America, East Asia and Other Asia confirm that most regions exhibit a linear trend in their cumulative hydrogen projects. One notable exception is East Asia, whose growth trajectory is almost hyperbolic.

Section \ref{sec:inconsistency} introduces frameworks that detect broad regional inconsistencies in the contribution of green vs fossil plants to the number and capacity of hydrogen projects over time. In particular, we demonstrate the growing inconsistency between differences in continents' proportion of green and non-green plants and capacity.

The future development of hydrogen plants must consider safety, profitability and, of course, the key aim of decarbonisation. While blue hydrogen plants discussed in this manuscript attempt to sequester their carbon emissions, this is imperfect and has significant risk \cite{IPCC}. Thus, we hope to see growth in the number of green hydrogen plants among all regions, and associated increases in their capacity. Monitoring these trends may be crucial to the future of decarbonisation as we drive toward net-zero.

\bibliographystyle{_elsarticle-num-names}
\bibliography{__newrefs.bib}
\biboptions{sort&compress}
\end{document}